\documentclass[11pt,a4paper]{article}
\usepackage{amssymb}
\usepackage{amsmath}
\usepackage{amsthm}

\begin{document}
\title{\Large On the MacWilliams Theorem over Codes and Lattices}
\author{Zhiyong Zheng$^{1,2,3}$, Fengxia Liu$^{*2,1,3}$ and Kun Tian$^{*1,2,3}$
\\ \small{$^{1}$Engineering Research Center of Ministry of Education for Financial Computing}
\\ \small{and Digital Engineering, Renmin University of China, Beijing, 100872, China}
\\ \small{$^{2}$The Great Bay University, Dongguan, 523830, China}
\\ \small{$^{3}$Henan Academy of Sciences, Zhengzhou, 450046, China}
\\ \small{$^{*}$Corresponding author email: shunliliu@gbu.edu.cn, tkun19891208@ruc.edu.cn}}

\date{}
\maketitle

\noindent\textbf{Abstract}\quad Analogies between codes and lattices have been extensively studied for the last decades, in this dictionary, the MacWilliams identity is the finite analog of the Jacobi-Poisson formula of the Theta function. Motivated by the random theory of lattices, the statistical significance of MacWilliams theorem is considered, indeed, MacWilliams distribution provides a finite analog of the classical Gauss distribution. In particular, the MacWilliams distribution over quotient space of a code is statistical closed with the uniform distribution.

In the respect of lattices, the analogy of MacWilliams identity associated with nu-function was conjectured by Sol\'{e} in 1995. We give an answer to this problem in positive.\\


\section{Introduction}

The MacWilliams theorem for linear codes with the Hamming metric \cite{11,12} establishes an identity that relates the weight enumerator of a code to the weight enumerators of its dual code. Various authors have extended this work in different directions. One direction involves generalizing the weight enumerators to include more than two variables, such as the Lee and complete weight enumerators, and extending the concept to codes defined over alphabets beyond finite fields. For instance, Wan \cite{21} provided a MacWilliams theorem for codes over Galois rings. Another generalization involves adapting the notion of weight to consider multiple codewords simultaneously, leading to the generalized Hamming weights by Wei \cite{22} and MacWilliams-type results for $m$-tuple support enumerators by Kl{\o}ve \cite{09}, Shiromoto \cite{18}, Simonis \cite{19}, and Ray-Chaudhuri and Siap \cite{15}. Britz \cite{04,05} further generalized some of these results and provided matroid-theoretic proofs. Additionally, Britz \cite{03} outlined new and extensive connections between weight enumerators and Tutte polynomials of matroids.

We first consider the MacWilliams theorem for effective length weight enumerators of an $m$-tuple of codes $C_1, C_2, \cdots, C_m$, which need not be identical. While similar results have been presented by Britz \cite{04}, Kaplan \cite{08}, and Shiromoto \cite{18}, our focus is on the statistical significance of the MacWilliams identities. We introduce new concepts for a code, such as the MacWilliams distribution and the smoothing parameter. In fact, we show that the MacWilliams distribution is a finite analog of the classical Gaussian distribution in $\mathbb{R}^n$. Specifically, we find that the MacWilliams distribution closely approximates the uniform distribution over the quotient space of a code.

Analogies between codes and lattices have been extensively studied over the past few decades. The classical problem of counting lattice points in Euclidean spheres involves the use of the Jacobi Theta function. In this context, the MacWilliams formula can be seen as the finite analog of the Poisson formula for the Jacobi Theta function. The more recent problem of counting lattice points in pyramids for the $L^1$-norm involves the nu-function of lattices \cite{02,20}. Both problems have significant applications in multidimensional vector quantization \cite{02,17}. In this paper, we present a new analog of the MacWilliams identity over lattices with the nu-function, a problem conjectured by Sol\'{e} \cite{25,20}. We provide a positive solution to this problem.

\subsection{$m$-Tuple MacWilliams Identity}

First, we give the necessary definitions to state the MacWilliams' original theorem \cite{12}. Let $F_q$ be a finite field of $q$ elements, $n$ a positive integer, and $C\subset F_q^n$ a linear code. Let $|C|$ be the number of codewords of $C$, $<a,b>$ be the usual pairing on $F_q^n$. The Hamming weight of any $x\in F_q^n$, denote by $w(x)$, is the number of nonzero coordinates of $x$. We denote by $W_C(z)$, the Hamming weight enumerator of $C$ in the indeterminates $z$:
\begin{equation*}
W_C(z)=\sum\limits_{x\in C} z^{w(x)},
\end{equation*}
or an equivalent homogeneous form in two indeterminates $z_1$ and $z_2$
\begin{equation*}
W_C(z_1,z_2)=\sum\limits_{x\in C} z_1^{w(x)}z_2^{n-w(x)}.
\end{equation*}
Let $C\subset F_q^n$ be any linear code and $C^{\bot}$ be its dual code, MacWilliams \cite{07} (or also see \cite{11}) showed that
\begin{equation*}
W_{C^{\bot}}(z)=\frac{1}{|C|}(1+(q-1)z)^n W_C(\frac{1-z}{1+(q-1)z}).  \tag{1.1}
\end{equation*}
Let $z=z_1/z_2$, one has the homogeneous form of MacWilliams identity immediately (see Theorem 9.32 of \cite{10})
\begin{equation*}
W_{C^{\bot}}(z_1,z_2)=\frac{1}{|C|} W_C(z_2-z_1,z_2+(q-1)z_1).  \tag{1.2}
\end{equation*}

Some authors consider more than one code at a time, this leads to the $m$-tuple MacWilliams identity as follows.

Let $C_1,C_2,\cdots,C_m \subset F_q^n$ be $m$ codes, not necessary  being the  same. Let $F_q^{m\times n}$ be the matrix ring of all $m\times n$ matrices over $F_q$. A matrix $x\in F_q^{m\times n}$, we denote by $ew(x)$, the effective length weight of $x$, is the number of nonzero columns of $x$. Let $\underline{C}=C_1\times C_2\times\cdots \times C_m$, $\underline{C}^{\bot}=C_1^{\bot}\times C_2^{\bot}\times \cdots \times C_m^{\bot}$, and $|\underline{C}|=|C_1|\times \cdots \times |C_m|$. We regard any an element $x=(x_1,x_2,\cdots,x_m)\in \underline{C}$ as a matrix of $F_q^{m\times n}$ by
\begin{equation*}
x=\begin{pmatrix} x_1 \\ x_2 \\ \vdots \\ x_m \end{pmatrix} \in F_q^{m\times n},\quad x_i\in C_i.
\end{equation*}
The effective length weight enumerator of $\underline{C}$ is given by
\begin{equation*}
W_{\underline{C}}^{(m)}(z)=\sum\limits_{x\in \underline{C}} z^{ew(x)}.
\end{equation*}
The following theorem is a generalization of MacWilliams identity for $m$ codes, which is a main result of \cite{08}.

\textbf{Theorem 1}\ \ Let $C_1,C_2,\cdots,C_m$ be any $m$ linear codes over $F_q^n$, then we have
\begin{equation*}
W_{{\underline{C}}^{\bot}}^{(m)}(z)=\frac{1}{|\underline{C}|} (1+(q^m-1)z)^n W_{\underline{C}}^{(m)}(\frac{1-z}{1+(q^m-1)z}).  \tag{1.3}
\end{equation*}

This result allows one to compare the effective length of $m$-tuple of vectors draw from different linear codes, and gives a generalization of an earlier work of Shiromoto \cite{18} concerning the effective length of $m$-tuple vectors from the same linear code. Taking $z=z_1/z_2$ in (1.3), one obtain the homogeneous form of the $m$-tuple MacWilliams identity, which is the Theorem 6 of \cite{08}. We use the finite Fourier transform over $F_q^{m\times n}$ to give an alternative proof. The proof is not only natural and brief, but also reflects the statistical significance of the MacWilliams theorem.

\subsection{MacWilliams Distribution}

The use of Gaussian distribution in the study of random lattices is a standard technique (see, for example, \cite{24}). In \cite{01,13,16}, a few authors used it to show that some important cryptographic results. In \cite{01}, for example, Gaussian distribution are used to prove that the certain lattice problems are in coNP. In \cite{13}, Micciancio and Regev introduced the smoothing parameter for a lattice, which plays a key role in proving of from worst-case to average-case reductions.

In this paper, we introduce the smoothing parameter for  code and show that the statistical significance of the MacWilliams theorem  provides a finite analog of Gaussian distribution over a code. To state our result, let $X$ and $Y$ be two discrete type random variables over a finite set $\Gamma$, we define the statistical distance between $X$ and $Y$ by $\Delta(X,Y)$ (see Definition 2.1, \cite{13})
\begin{equation*}
\Delta(X,Y)=\frac{1}{2} \sum\limits_{a\in \Gamma}|\text{Pr}\{X=a\}-\text{Pr}\{Y=a\}|.
\end{equation*}
If the statistical distance $\Delta(X,Y)$ can be arbitrary small, we call $X$ and $Y$ are statistical closed on $\Gamma$.

Let $C\subset F_q^n$ be a code of length $n$, $z\in (0,1)$  a real number, we define the MacWilliams distribution as follows
\begin{equation*}
p(x)=z^{w(x)}/\sum\limits_{x\in C} z^{w(x)},\quad x\in C.
\end{equation*}
Since $\sum\limits_{x\in C} p(x)=1$, which corresponds a discrete type random variable taking value on $C$. Let $F_q^n/C$ be the quotient space, we define by $D_C$, a discrete random variable over $F_q^n/C$, its probability distribution function given by
\begin{equation*}
g(x)=\sum\limits_{c\in C}z^{w(x+c)}/(1+(q-1)z)^n,\quad x\in F_q^n/C.
\end{equation*}
We note that if $C$ is a linear code, then
\begin{equation*}
\sum\limits_{x\in F_q^n/C} g(x)=1.
\end{equation*}

\textbf{Definition 1.1}\ \ Let $\epsilon>0$ be a positive real number, $C\subset F_q^n$ be a code and $C^{\bot}$ be its dual code, the smoothing parameter $\eta_{\epsilon}(C)$ given by
\begin{equation*}
\eta_{\epsilon}(C)=\min\{z \big| \ 0<z<1, \ \sum\limits_{x\in C^{\bot}\backslash \{0\}} (\frac{1-z}{1+(q-1)z})^{w(x)}<\epsilon \}.  \tag{1.4}
\end{equation*}

One of the main result of this paper is to show  the following theorem on MacWilliams distribution.

\textbf{Theorem 2}\ \ Let $C\subset F_q^n$ be a linear code, then the random variable $D_C$ over $F_q^n/C$ is statistical closed with the uniform distribution over $F_q^n/C$.

This result shows that the MacWilliams theorem provides a finite analog of the classical Gaussian distribution in $\mathbb{R}^n$.

\subsection{Theta Function and nu-Function over Lattices}

Let $\Lambda\subset \mathbb{R}^n$ be a lattice of rank $n$, the Theta function over $\Lambda$ is defined as the formal power series:
\begin{equation*}
\theta_{\Lambda}(q)=\sum\limits_{x\in \Lambda} q^{|x|^2},
\end{equation*}
where $q=e^{\pi iz}$, $|x|^2=x_1^2+\cdots+x_n^2$ is the common  Euclidean norm. It is easily seen that the coefficient $N_m$ of $q^{m^2}$ in $\theta_{\Lambda}(q)$ counts the number of lattice points at distance $m$ from the origin in $\mathbb{R}^n$. The classical Jacobi Theta function is
\begin{equation*}
\Theta(\xi|z)=\sum\limits_{m=-\infty}^{+\infty} e^{2mi\xi+\pi izm^2}.
\end{equation*}
Some special examples of Theta function given by $\theta_3(q)=\Theta(0|z)$, which is the Theta function over lattice $\mathbb{Z}$ and $\theta_2(q)=e^{\pi iz/4} \Theta(\frac{\pi z}{2}|z)$, which is the Theta function over lattice $\mathbb{Z}$ to co-sets, and it is easy to see that $\theta_2(q^2)=\theta_{2\mathbb{Z}+1}(q)$.

An analog of MacWilliams theorem over lattices is the following Jacobi-Poisson formula for Theta function (see \cite{20}, or proposition 2.1 of \cite{07})
\begin{equation*}
\theta_{\Lambda^*}(e^{\pi iz})=\text{det}(\Lambda)(\frac{i}{z})^{\frac{\pi}{2}}\theta_{\Lambda}(e^{-\frac{i\pi}{z}}),  \tag{1.5}
\end{equation*}
where $\Lambda^*$ is the dual lattice of $\Lambda$, and $\text{det}(\Lambda)$ is the determinant of $\Lambda$.

To explain why the above Poisson formula is a generalization of MacWill-iams theorem to lattice, we need some connections between codes and lattices, the most simple way to associate a lattice with a code is the following construction \cite{20}.

\hspace{0.5cm} \textbf{Construction A:}

Let $C\subset F_q^n$ be a binary code coordinatized w.r.t a special basis of $F_2^n$. A lattice $A(C)$ is constructed by
\begin{equation*}
A(C)=\{(x_1,x_2,\cdots,x_n)\in \mathbb{Z}^n\ \big|\ \exists c\in C\ \text{such that}\ x\equiv c\ (\text{mod}\ 2)\}.  \tag{1.6}
\end{equation*}
It is known that \cite{20}
\begin{equation*}
\theta_{A(C)}=W_C(\theta_3(q^4),\theta_2(q^4)),  \tag{1.7}
\end{equation*}
where $W_C(z_1,z_2)$ is the weight enumerator of $C$. Indeed, one could derive the MacWilliams formula over $C$ from the Jacobi-Poisson formula (1.5) by using construction A and the relation that $A(C)^*=\frac{1}{2}A(C^{\bot})$ ((3) of \cite{20}, or lemma 4.3 below).

Motivated by vector quantizing application \cite{02}, some authors \cite{02,17,20,23} replace the $L^2$-norm by $L^1$-norm, this leads to hyperbolic trigonometric function appeared, the nu-function over a lattice $\Lambda$ is defined by
\begin{equation*}
\nu_{\Lambda}(z)=\sum\limits_{x\in \Lambda}z^{|x|_1}=\sum\limits_{n=0}^{+\infty} z^n |\{x\in \Lambda: |x|_1=n\}|,
\end{equation*}
where $|x|_1=\sum\limits_{i=1}^n |x_i|$ is the $L^1$-norm. Although the Poisson formula for nu-function is missing, the same analogy of MacWilliams identity exists. It was conjectured by \cite{20} that
\begin{equation*}
2^{\frac{n}{2}} \nu_{\Lambda^*}(\text{tanh}^2 (\frac{\beta}{2}))=\text{det}(\Lambda)(\text{sinh}(2\beta))^{\frac{n}{2}}\nu_{\Lambda}(\text{tanh}(\frac{\alpha}{2})),  \tag{1.8}
\end{equation*}
where $\Lambda=A(C)$ associated with a binary linear code $C$, and parameters $\alpha$ and $\beta$ be connected by the relation $e^{-2\beta}=\text{tanh}(\alpha)$.

The main purpose of this paper is to show that this conjecture, we give a positive answer to this problem.

The paper is organized as follows. In section 2, we develop the finite Fourier transform technique and give a brief proof to Theorem 1. In section 3, we give a proof to Theorem 2 by using MacWilliams identity. We state and prove Theorem 3 in section 4, which gives an answer to conjecture 1 of Sol\'{e} \cite{20}. Finally, we give a conclusion in section 5, a few new consideration are presented.

\section{Finite Fourier Transform over the Matrix Ring}

Let $F_q$ be a finite field of $q$ elements, $\psi$ be a non-trivial additive character of $F_q$. Suppose that $f$ is a function defined over $F_q^n$, the finite Fourier transform FTf (sometimes called Hadamard transform) given by
\begin{equation*}
FTf(x)=\sum\limits_{\xi\in F_q^n} f(\xi)\psi(<x,\xi>),\quad x\in F_q^n,
\end{equation*}
where $<x,\xi>$ is the usual pairing.

First, we generalize this transform to a matrix ring. Let $F_q^{m\times n}$ be the ring consisting of all $m\times n$ matrices over $F_q$. Any two matrices $x,y\in F_q^{m\times n}$, then $x^T y$ is an $n\times n$ square matrix. Let Tr$(x^T y)$ be the trace of $x^T y$, then Tr$(x^T y)\in F_q$. We denote by $<x,y>=\text{Tr}(x^T y)$, the pairing of $x$ and $y$.

\textbf{Definition 2.1}\ \ Suppose that $f$ is a function defined over $F_q^{m\times n}$, the finite Fourier transform FTf of $f$ is defined by
\begin{equation*}
FTf(x)=\sum\limits_{\xi\in F_q^{m\times n}} f(\xi)\psi(<x,\xi>),\quad x\in F_q^{m\times n}.
\end{equation*}
It is easy to see that
\begin{equation*}
f(x)=\frac{1}{q^{mn}} \sum\limits_{\xi\in F_q^{m\times n}} FTf(\xi) \psi(-<x,\xi>),\quad x\in F_q^{m\times n}.
\end{equation*}
Let $C_1,C_2,\cdots,C_m$ be $m$ codes over $F_q$ of length $n$, $\underline{C}=C_1\times \cdots \times C_m$, we regard $\underline{C}$ as a subset of $F_q^{m\times n}$ according to $x=(x_1,\cdots,x_m)^T$, where $x_i\in C_i$. Let $\chi_{\underline{C}}$ be the characteristic function of $\underline{C}$, that means $\chi_{\underline{C}}(x)=1$, if $x\in \underline{C}$, and $\chi_{\underline{C}}(x)=0$, if $x\notin \underline{C}$.

\textbf{Lemma 2.1}\ \ If $C_i$ ($1\leqslant i\leqslant m$) is the linear code of $F_q^n$, then we have
\begin{equation*}
\chi_{\underline{C}^{\bot}}(x)=\frac{1}{|\underline{C}|}\text{FT}\chi_{\underline{C}}(x),
\end{equation*}
where $\underline{C}^{\bot}=C_1^{\bot}\times C_2^{\bot}\times\cdots \times C_m^{\bot}$.

\textbf{Proof:} By the definition of FT$\chi_{\underline{C}}$, we see that for any $x\in F_q^{m\times n}$,
\begin{equation*}
\text{FT}\chi_{\underline{C}}(x)=\sum\limits_{\xi\in F_q^{m\times n}} \chi_{\underline{C}}(\xi)\psi(<x,\xi>)=\sum\limits_{\xi\in \underline{C}}\psi(<x,\xi>).
\end{equation*}
If we write
\begin{equation*}
x=\begin{pmatrix} x_1\\ \vdots\\ x_m \end{pmatrix},\ x_i\in F_q^n,\ \text{and}\ \xi=\begin{pmatrix} \xi_1\\ \vdots\\ \xi_m \end{pmatrix},\ \xi_i\in C_i,
\end{equation*}
then
\begin{equation*}
<x,\xi>=\text{Tr}(x^T \xi)=\sum\limits_{i=1}^m <x_i,\xi_i>.  \tag{2.1}
\end{equation*}
It follows that
\begin{equation*}
\text{FT}\chi_{\underline{C}}(x)=\mathop{\prod}\limits_{i=1}^m \sum\limits_{\xi_i\in C_i} \psi(<x_i,\xi_i>).
\end{equation*}
It is known that (see, for example, lemma 9.31 of \cite{10})
\begin{equation*}
\sum\limits_{\xi_i\in C_i} \psi(<x_i,\xi_i>)=\left\{\begin{array}{cc} |C_i|, & \text{if}\ x_i\in C_i^{\bot}. \\ 0, & \text{otherwise.} \end{array}\right.
\end{equation*}
We have FT$\chi_{\underline{C}}(x)=|\underline{C}|\chi_{\underline{C}^{\bot}}(x)$.\\
\hspace*{12cm} $\Box$

Using the above observation, we have the following more general Poisson formula over matrix ring.

\textbf{Lemma 2.2}\ \ Let $\underline{C}\subset F_q^{m\times n}$ be the block linear codes over $F_q$, and $f(x)$ be arbitrary function over $F_q^{m\times n}$, then
\begin{equation*}
\sum\limits_{x\in \underline{C}^{\bot}}f(x)=\frac{1}{|\underline{C}|} \sum\limits_{x\in \underline{C}}\text{FT}f(x).
\end{equation*}

\textbf{Proof:}
\begin{equation*}
\sum\limits_{x\in \underline{C}^{\bot}}f(x)=\sum\limits_{x\in F_q^{m\times n}} \chi_{\underline{C}^{\bot}}(x)f(x)
\end{equation*}
\begin{equation*}
\qquad\qquad\qquad\quad=\frac{1}{|\underline{C}|} \sum\limits_{x\in F_q^{m\times n}} f(x)\text{FT}\chi_{\underline{C}}(x)
\end{equation*}
\begin{equation*}
\qquad\qquad\qquad\qquad\qquad\qquad\qquad=\frac{1}{|\underline{C}|} \sum\limits_{\xi\in F_q^{m\times n}} \chi_{\underline{C}}(\xi) \sum\limits_{x\in F_q^{m\times n}} f(x)\psi(<x,\xi>)
\end{equation*}
\begin{equation*}
\qquad\quad=\frac{1}{|\underline{C}|} \sum\limits_{\xi\in \underline{C}}\text{FT}f(\xi).
\end{equation*}
\hspace*{12cm} $\Box$

Now, we give a brief proof to Theorem 1.

\textbf{Proof of Theorem 1:}

Let $f(x)=z^{ew(x)}$, where $x\in F_q^{m\times n}$, and $ew(x)$ is the effective length weight function given by section 1. To prove Theorem 1, by the Poisson formula, we only show that
\begin{equation*}
\text{FT}z^{ew(x)}=(1+(q^m-1)z)^n (\frac{1-z}{1+(q^m-1)z})^{ew(x)}.  \tag{2.2}
\end{equation*}

We write $x=(x_1,x_2,\cdots,x_n)\in F_q^{m\times n}$, $x_i\in F_q^m$, and $\xi=(\xi_1,\xi_2,\cdots,\xi_n)$ $\in F_q^{m\times n}$, $\xi_i\in F_q^m$, then (comparing with (2.1))
\begin{equation*}
<x,\xi>=\text{Tr}(x^T \xi)=\sum\limits_{i=1}^n <\xi_i,x_i>.  \tag{2.3}
\end{equation*}
It follows that
\begin{equation*}
\text{FT}f(x)=\sum\limits_{\xi\in F_q^{m\times n}} f(\xi) \psi(<x,\xi>)
\end{equation*}
\begin{equation*}
\qquad\qquad\ =\sum\limits_{\xi\in F_q^{m\times n}} z^{ew(\xi)} \psi(<x,\xi>)
\end{equation*}
\begin{equation*}
\qquad\qquad\qquad\qquad\ \ \ =\sum\limits_{\xi\in F_q^{m\times n}} z^{ew(\xi_1)+\cdots+ew(\xi_n)} \psi(<x,\xi>)
\end{equation*}
where $ew(\xi_i)=1$, if $\xi_i\neq 0$, and $ew(\xi_i)=0$, if $\xi_i=0$ is a zero vector. Therefore, we have
\begin{equation*}
\text{FT}f(x)=\mathop{\prod}\limits_{i=1}^n (\sum\limits_{\xi_i\in F_q^m} z^{ew(\xi_i)} \psi(<x_i,\xi_i>))
\end{equation*}
\begin{equation*}
\qquad\qquad\quad=\mathop{\prod}\limits_{i=1}^n (1+z\sum\limits_{\xi_i\in F_q^m \backslash \{0\}} \psi(<x_i,\xi_i>)).
\end{equation*}
It is known that
\begin{equation*}
\sum\limits_{\xi_i\in F_q^m \backslash \{0\}} \psi(<x_i,\xi_i>)=\left\{\begin{array}{cc} q^m-1, & \text{if}\ x_i=0. \\ -1, & \text{if}\ x_i\neq 0. \end{array}\right.
\end{equation*}
It follows that
\begin{equation*}
\text{FT}z^{ew(x)}=\mathop{\prod}\limits_{i=1}^n (1+z\sum\limits_{\xi_i\in F_q^m \backslash \{0\}} \psi(<x_i,\xi_i>))\qquad
\end{equation*}
\begin{equation*}
\qquad\qquad\quad\ =\mathop{\prod}\limits_{i=1}^n ((1-z)^{ew(x_i)}(1+(q^m-1)z)^{1-ew(x_i)})
\end{equation*}
\begin{equation*}
\qquad\qquad\ =(1+(q^m-1)z)^n (\frac{1-z}{1+(q^m-1)z})^{ew(x)}.  \tag{2.4}
\end{equation*}
We complete the proof of Theorem 1.\\
\hspace*{12cm} $\Box$

In the proof  of Theorem 1, it is important that the function $z^{ew(x)}$ is almost fixed under the finite Fourier transform over $F_q^{m\times n}$, thus this function provides a finite analog of Gaussian measure, and we obtain a finite analog of Gaussian distribution by using the fixed point. The statistical significance of MacWilliams theorem is revealed by this observation.

\section{Proof of Theorem 2}

Let $C\subset F_q^n$ be a linear code of length $n$, $C^{\bot}$ be its dual code, $0<z<1$ be a real number parameter. The smoothing parameter $\eta_{\epsilon}(C)$ is defined as (see Definition 1.1)
\begin{equation*}
\eta_{\epsilon}(C)=\min\{z\ \big|\ 0<z<1, \sum\limits_{x\in C^{\bot}\backslash \{0\}} (\frac{1-z}{1+(q-1)z})^{w(x)}<\epsilon \}.  \tag{3.1}
\end{equation*}

If we take $C=0$, a zero code in MacWilliams identity, it is easy to see that
\begin{equation*}
\sum\limits_{x\in F_q^n} z^{w(x)}=(1+(q-1)z)^n.
\end{equation*}

The following probability distribution function corresponds a discrete random variable $D_C$ over the quotient space $F_q^n/C$
\begin{equation*}
g(x)=\sum\limits_{c\in C} z^{w(x+c)}/(1+(q-1)z)^n,\quad x\in F_q^n/C.
\end{equation*}

To show that $D_C$ is statistical closed to the uniform distribution $U(F_q^n/C)$, we only prove the following proposition.

\textbf{Proposition 3.1}\ \ Let $C\subset F_q^n$ be a linear code, then for arbitrary $\epsilon>0$ and $\eta_{\epsilon}(C)\leqslant z<1$, the statistical distance $D_C$ and the uniform distribution $U(F_q^n/C)$ satisfies the following inequality:
\begin{equation*}
\Delta(D_C,U(F_q^n/C))\leqslant \frac{1}{2}\epsilon.
\end{equation*}

\textbf{Proof:} Let $f(x)$ be any a function over $F_q^n$, by lemma 2.2 we have the Poisson formula over code $C$,
\begin{equation*}
\sum\limits_{x\in C^{\bot}} f(x)=\frac{1}{|C|} \sum\limits_{x\in C} \text{FT}f(x).  \tag{3.2}
\end{equation*}
If $x\in F_q^n/C$ and $z$ are given, let
\begin{equation*}
f(c)=z^{w(x+c)}/(1+(q-1)z)^n,\quad \forall c\in C.
\end{equation*}
The probability distribution function $g(x)$ of $D_C$ is
\begin{equation*}
g(x)=\sum\limits_{c\in C} f(c),\quad \forall x\in F_q^n/C.
\end{equation*}
First, we calculate the Fourier transform FT$f(c)$. By definition,
\begin{equation*}
\text{FT}f(c)=\sum\limits_{\xi\in F_q^n} f(\xi)\psi(<\xi,c>)\qquad\qquad\qquad\qquad
\end{equation*}
\begin{equation*}
\qquad\qquad=\frac{1}{(1+(q-1)z)^n}\sum\limits_{\xi\in F_q^n} z^{w(x+\xi)} \psi(<\xi,c>).
\end{equation*}
Let $x+\xi=\xi_1$. By (2.4) we have ($m=1$ in (2.4))
\begin{equation*}
\text{FT}f(c)=\frac{\psi(-<x,c>)}{(1+(q-1)z)^n}\sum\limits_{\xi_1\in F_q^n} z^{w(\xi_1)} \psi(<\xi_1,c>)
\end{equation*}
\begin{equation*}
\ \ =(\frac{1-z}{1+(q-1)z})^{w(c)}\psi(-<x,c>).
\end{equation*}
According to the Poisson formula (3.2), it follows that
\begin{equation*}
g(x)=\sum\limits_{c\in C} f(c)=\frac{1}{|C^{\bot}|}\sum\limits_{c\in C^{\bot}}\text{FT}f(c)\qquad\quad
\end{equation*}
\begin{equation*}
\qquad\qquad\quad\ =\frac{1}{|C^{\bot}|}\sum\limits_{c\in C^{\bot}}(\frac{1-z}{1+(q-1)z})^{w(c)}\psi(-<x,c>).
\end{equation*}
Therefore, the statistical distance $\Delta(D_C,U(F_q^n/C))$ may be calculated by
\begin{equation*}
\Delta(D_C,U(F_q^n/C))=\frac{1}{2} \sum\limits_{x\in F_q^n/C} \Big|g(x)-\frac{1}{|C^{\bot}|}\Big|
\end{equation*}
\begin{equation*}
=\frac{1}{2|C^{\bot}|} \sum\limits_{x\in F_q^n/C} \Big|\sum\limits_{c\in C^{\bot}} (\frac{1-z}{1+(q-1)z})^{w(c)} \psi(-<x,c>)-1\Big|
\end{equation*}
\begin{equation*}
=\frac{1}{2|C^{\bot}|} \sum\limits_{x\in F_q^n/C} \Big|\sum\limits_{c\in C^{\bot} \backslash \{0\}} (\frac{1-z}{1+(q-1)z})^{w(c)} \psi(-<x,c>)\Big|
\end{equation*}
\begin{equation*}
\leqslant\frac{1}{2} \Big|\sum\limits_{c\in C^{\bot} \backslash \{0\}} (\frac{1-z}{1+(q-1)z})^{w(c)}\Big|.\qquad\qquad\qquad\qquad\qquad\ \ \
\end{equation*}
If $\eta_{\epsilon}(C)\leqslant z<1$, we have
\begin{equation*}
\Big|\sum\limits_{c\in C^{\bot} \backslash \{0\}} (\frac{1-z}{1+(q-1)z})^{w(c)}\Big|\leqslant \epsilon.
\end{equation*}
It follows that
\begin{equation*}
\Delta(D_C,U(F_q^n/C))\leqslant \frac{1}{2}\epsilon.
\end{equation*}
We complete the proof of proposition 3.1.\\
\hspace*{12cm} $\Box$

\section{The nu-Function over Lattices}

Let $\Lambda\subset \mathbb{R}^n$ be a lattice of rank $n$, the nu-function over $\Lambda$ is defined by
\begin{equation*}
\nu_{\Lambda}(z)=\sum\limits_{x\in \Lambda}z^{|x|_1},
\end{equation*}
where $|x|_1=|x_1|+|x_2|+\cdots+|x_n|$ is the $L^1$-norm. A function $f$ defined over $\mathbb{R}^n$, the Fourier transform given by
\begin{equation*}
\text{FT}f(x)=\int_{\mathbb{R}^n} f(\xi)e^{-2\pi i<x,\xi>}\mathrm{d}\xi,\quad x\in \mathbb{R}^n,
\end{equation*}
where $<x,\xi>=\sum\limits_{i=1}^n x_i \xi_i$ is the usual pairing of $x$ and $\xi$. It is easy to see that the Gauss function $e^{-\pi |x|^2}$ is a fixed point under the Fourier transform, or slightly more general, with a parameter $s$, one has
\begin{equation*}
\text{FT}e^{-\pi |x/s|^2}=s^n e^{-\pi |sx|^2}.
\end{equation*}
Indeed, the MacWilliams weight enumerator $z^{w(x)}$ is a finite analog of Gauss function. The following Poisson formula over lattices is classical (see Theorem 2.3 of \cite{07}, or lemma 1.9 of \cite{24}).

\textbf{Lemma 4.1}\ \ Let $\Lambda \subset \mathbb{R}^n$ be a lattice of rank $n$, $f: \mathbb{R}^n \rightarrow \mathbb{C}$ be a function which satisfies the following conditions (V1), (V2), (V3):

(V1) $\int_{\mathbb{R}^n}|f(x)|\mathrm{d}x<\infty$.

(V2) The series $\sum\limits_{x\in \Lambda} |f(x+u)|$ converges uniformly for all $u$ belonging to a compact subset of $\mathbb{R}^n$.

(V3) The series $\sum\limits_{x\in \Lambda^*} \text{FT}f(x)$ is absolutely convergent.\\
Then we have
\begin{equation*}
\sum\limits_{x\in \Lambda}f(x)=\frac{1}{\text{det}(\Lambda)}\sum\limits_{x\in \Lambda^*}\text{FT}f(x).
\end{equation*}

Taking $f(x)=e^{\pi iz |x|^2}$ in lemma 4.1, we have the Jacobi-Poisson formula (see (1.5)) immediately. If one replace the $L^2$-norm by $L^1$-norm, and consider the nu-function over $\Lambda$, since $z^{|x|_1}$ is not a fixed point under the Fourier transform, the Poisson formula for nu-function over lattice $\Lambda$ is missing.

For some special lattices, for example $A(C)$, the lattices associated with the binary code $C$ by using construction A, the analogy of MacWilliams theorem for nu-function still exists. We state our main result as follows.

\textbf{Theorem 3}\ \ Let $\Lambda=A(C)$, $\alpha$ and $\beta$ be the parameters such that $e^{-2\beta}=\tanh(\alpha)$, then
\begin{equation*}
2^{\frac{n}{2}}\nu_{\Lambda^*}(\tanh^2(\frac{\beta}{2}))=\text{det}(\Lambda)(\sinh(2\beta))^{\frac{n}{2}}\nu_{\Lambda}(\tanh(\frac{\alpha}{2})),
\end{equation*}
where $\tanh$ and $\sinh$ are hyperbolic trigonometric functions.

This result is an open problem appeared in Sol\'{e} \cite{20} as the Conjecture 1, but in his original statement the relation $e^{-2\beta}=\tanh(\alpha)$ is missed by $e^{-\beta}=\tanh(\alpha)$. Of course, that is a print error, because he already noted the symmetric relation $e^{-2\alpha}=\tanh(\beta)$. To show that Theorem 3, we first prove a few auxiliary lemmas.

\textbf{Lemma 4.2}\ \ Let $L\subset \mathbb{R}^n$ be any a lattice, $z$ is a given parameter, then $\nu_{2L}(z)=\nu_{L}(z^2)$.

\textbf{Proof:} By the definition of nu-function, we have
\begin{equation*}
\nu_{2L}(z)=\sum\limits_{x\in 2L} z^{|x|_1}=\sum\limits_{y\in L} z^{|2y|_1}=\sum\limits_{y\in L} (z^2)^{|y|_1}=\nu_L(z^2).
\end{equation*}
\hspace*{12cm} $\Box$

\textbf{Lemma 4.3}\ \ Let $A(C)$ be a lattice associated with the binary code $C$, then the dual lattice of $A(C)$ given by
\begin{equation*}
A(C)^*=\frac{1}{2} A(C^{\bot}),
\end{equation*}
where $C^{\bot}$ is the dual code of $C$.

\textbf{Proof:} We first prove that $A(C)^*\subset \frac{1}{2}A(C^{\bot})$. For any $\alpha\in A(C)^*$, we note that $x\in A(C)$ whence $x\in C$, it follows that $<\alpha,x>\in \mathbb{Z}$ for all $x\in C$, and
\begin{equation*}
<2\alpha,x>\equiv 0\ (\text{mod}\ 2),\quad \forall x\in C.
\end{equation*}
This means that $2\alpha\ \text{mod}\ 2\in C^{\bot}$, and $2\alpha\in A(C^{\bot})$, which implies that $\alpha \in \frac{1}{2} A(C^{\bot})$.

To show that $\frac{1}{2}A(C^{\bot}) \subset A(C)^*$, for any $\beta\in \frac{1}{2}A(C^{\bot})$, or $2\beta\in A(C^{\bot})$, we see that $\beta\in A(C)^*$. Let $x\in C$, we have $<2\beta\ \text{mod}\ 2,x>=0$, which implies that $<2\beta,x>\equiv 0\ (\text{mod}\ 2)$. Thus we have $<\beta,x>\in \mathbb{Z}$ for all $x\in C$. Let $y\in A(C)$, denote $y\ \text{mod}\ 2=x_0\in C$, then $<\beta,y>\in \mathbb{Z}$ by $<\beta,x_0>\in \mathbb{Z}$, this leads to $\beta\in A(C)^*$, and $\frac{1}{2}A(C^{\bot}) \subset A(C)^*$. We have lemma 4.3.\\
\hspace*{12cm} $\Box$

\textbf{Lemma 4.4}\ \ Let $C\subset F_2^n$ be a binary code of length $n$, $\alpha$ be any a parameter, then we have
\begin{equation*}
\nu_{A(C)}(\tanh\frac{\alpha}{2})=W_C(\cosh\alpha,\sinh\alpha),
\end{equation*}
where $W_C(z_1,z_2)$ is the weight enumerator in the homogeneous form.

\textbf{Proof:} For any $x\in C$, we define a set $A(x)$ by
\begin{equation*}
A(x)=\{y\in \mathbb{Z}^n\ |\ y\equiv x\ (\text{mod}\ 2)\}.
\end{equation*}
It is easy to see that $A(x_1) \cap A(x_2)=\emptyset$, if $x_1\neq x_2$, so we have
\begin{equation*}
A(C)=\mathop{\cup}\limits_{x\in C} A(x),\ \text{and}\ \nu_{A(C)}(z)=\sum\limits_{x\in C}\nu_{A(x)}(z).
\end{equation*}
Let $x=(x_1,x_2,\cdots,x_n)$, then $A(x)=(x_1+2\mathbb{Z})\times (x_2+2\mathbb{Z})\times \cdots \times (x_n+2\mathbb{Z})$. It is not difficult to see
\begin{equation*}
\nu_{A(x)}(z)=\sum\limits_{y\in A(x)}z^{|y|_1}=(\sum\limits_{y'\in 2\mathbb{Z}}z^{|y'|})^{n-w(x)} (\sum\limits_{y''\in 2\mathbb{Z}+1}z^{|y''|})^{w(x)}
\end{equation*}
\begin{equation*}
=(\nu_{2\mathbb{Z}}(z))^{n-w(x)} (\nu_{2\mathbb{Z}+1}(z))^{w(x)}.\qquad\qquad\
\end{equation*}
It follows that
\begin{equation*}
\nu_{A(c)}(z)=\sum\limits_{x\in C} \nu_{A(x)}(z)=\sum\limits_{x\in C} (\nu_{2\mathbb{Z}}(z))^{n-w(x)} (\nu_{2\mathbb{Z}+1}(z))^{w(x)}
\end{equation*}
\begin{equation*}
=W_C(\nu_{2\mathbb{Z}}(z),\nu_{2\mathbb{Z}+1}(z)).\qquad\qquad\qquad\quad\
\end{equation*}
If the parameter $z$ satisfies $|z|<1$, one has
\begin{equation*}
\nu_{2\mathbb{Z}}(z)=\frac{1+z^2}{1-z^2},\ \text{and}\ \nu_{2\mathbb{Z}+1}(z)=\frac{2z}{1-z^2}.
\end{equation*}
By the above calculate, we obtain
\begin{equation*}
\nu_{A(C)}(z)=W_C(\frac{1+z^2}{1-z^2},\frac{2z}{1-z^2}).
\end{equation*}
Let $\alpha$ be a parameter such that $z=\tanh \frac{\alpha}{2}$, it is easily seen that
\begin{equation*}
\frac{1+z^2}{1-z^2}=\cosh\alpha,\ \text{and}\ \frac{2z}{1-z^2}=\sinh\alpha.
\end{equation*}
We have
\begin{equation*}
\nu_{A(C)}(\tanh\frac{\alpha}{2})=W_C(\cosh\alpha,\sinh\alpha).
\end{equation*}
This is the proof of lemma 4.4.\\
\hspace*{12cm} $\Box$

\textbf{Lemma 4.5}\ \ For any binary code $C\subset F_2^n$, we have
\begin{equation*}
|C|\text{det}(A(C))=2^n,
\end{equation*}
where $A(C)$ is the lattice associated with $C$, and $\text{det}(A(C))$ is the determinant of $A(C)$.

\textbf{Proof:} Considering the additive group homomorphism $\sigma: \mathbb{Z}^n \rightarrow F_2^n$ given by
\begin{equation*}
(x_1,x_2,\cdots,x_n)\in \mathbb{Z}^n \xrightarrow[]{\ \ \sigma\ \ } (x_1\ \text{mod}\ 2, x_2\ \text{mod}\ 2, \cdots, x_n\ \text{mod}\ 2)\in F_2^n.
\end{equation*}
Obviously, $\sigma^{-1}(C)=A(C)$, one has the following quotient group isomorphism
\begin{equation*}
\mathbb{Z}^n/A(C)\cong F_2^n/C.
\end{equation*}
It is known that
\begin{equation*}
\text{det}(A(C))=|\mathbb{Z}^n/A(C)|.
\end{equation*}
Thus, we have
\begin{equation*}
|C|\text{det}(A(C))=|F_2^n|=2^n.
\end{equation*}
\hspace*{12cm} $\Box$

\textbf{Lemma 4.6}\ \ For any binary code $C\subset F_2^n$, and two parameters $\alpha$ and $\beta$ satisfying $e^{-2\beta}=\tanh\alpha$, then we have
\begin{equation*}
\text{det}(A(C))(\frac{\sinh 2\beta}{2})^{\frac{n}{2}} (\frac{\cosh\alpha}{\cosh\beta})^n=\frac{1}{|C|}(1+\tanh\beta)^n.
\end{equation*}
where $\text{det}(A(C))$ is the determinant of $A(C)$.

\textbf{Proof:} Based on the equality $e^{-2\beta}=\tanh\alpha$ we can get $e^{-2\alpha}=\tanh\beta$ symmetrically, then
\begin{equation*}
\cosh^2\alpha=\frac{(e^{\alpha}+e^{-\alpha})^2}{4}=\frac{e^{2\alpha}+e^{-2\alpha}+2}{4}=(\frac{\cosh \beta}{\sinh \beta}+\frac{\sinh \beta}{\cosh \beta}+2)/4
\end{equation*}
\begin{equation*}
=\frac{(\cosh\beta+\sinh\beta)^2}{4\sinh\beta \cosh\beta}=\frac{(\cosh\beta+\sinh\beta)^2}{2\sinh 2\beta}.\qquad\qquad
\end{equation*}
So we have
\begin{equation*}
2\sinh 2\beta \cosh^2\alpha=(\cosh\beta+\sinh\beta)^2
\end{equation*}
and
\begin{equation*}
2\sinh 2\beta (\frac{\cosh\alpha}{\cosh\beta})^2=(\frac{\cosh\beta+\sinh\beta}{\cosh\beta})^2=(1+\tanh\beta)^2.
\end{equation*}
Take the $\frac{n}{2}$ power of both sides in the above equality, we get
\begin{equation*}
2^{\frac{n}{2}}(\sinh 2\beta)^{\frac{n}{2}} (\frac{\cosh\alpha}{\cosh\beta})^n=(1+\tanh\beta)^n.
\end{equation*}
By lemma 4.5, $\text{det}(A(C))|C|=2^n$, therefore,
\begin{equation*}
\text{det}(A(C))|C|2^{-\frac{n}{2}}(\sinh 2\beta)^{\frac{n}{2}} (\frac{\cosh\alpha}{\cosh\beta})^n=(1+\tanh\beta)^n,
\end{equation*}
which is equivalent to
\begin{equation*}
\text{det}(A(C))(\frac{\sinh 2\beta}{2})^{\frac{n}{2}} (\frac{\cosh\alpha}{\cosh\beta})^n=\frac{1}{|C|}(1+\tanh\beta)^n.
\end{equation*}
Then we finish the proof of this lemma.\\
\hspace*{12cm} $\Box$

Now, we give a proof of Theorem 3.

\textbf{Proof of Theorem 3:}

Let $C\subset F_2^n$ be arbitrary a binary code of length $n$, $A(C)$ be the lattice associated with $C$. By lemma 4.2, we have
\begin{equation*}
\nu_{A(C)^*}(\tanh^2 \frac{\beta}{2})=\nu_{2A(C)^*}(\tanh \frac{\beta}{2}).  \tag{4.1}
\end{equation*}
By lemma 4.3, we have
\begin{equation*}
\nu_{2A(C)^*}(\tanh \frac{\beta}{2})=\nu_{A(C^{\bot})}(\tanh \frac{\beta}{2})  \tag{4.2}
\end{equation*}
and by lemma 4.4
\begin{equation*}
\nu_{A(C^{\bot})}(\tanh \frac{\beta}{2})=W_{C^{\bot}}(\cosh\beta,\sinh\beta).  \tag{4.3}
\end{equation*}
Combining with (4.1), (4.2) and (4.3), we have
\begin{equation*}
\nu_{A(C)^*}(\tanh^2 \frac{\beta}{2})=W_{C^{\bot}}(\cosh\beta,\sinh\beta).  \tag{4.4}
\end{equation*}
Similarly, in (4.3) we replace $C^{\bot}$ by $C$, which gives
\begin{equation*}
\nu_{A(C)}(\tanh \frac{\beta}{2})=W_{C}(\cosh\beta,\sinh\beta).
\end{equation*}
Therefore, to prove Theorem 3, it is sufficient to prove
\begin{equation*}
W_{C^{\bot}}(\cosh\beta,\sinh\beta)=\text{det}(A(C))(\frac{\sinh 2\beta}{2})^{\frac{n}{2}} W_C(\cosh\alpha,\sinh\alpha).  \tag{4.5}
\end{equation*}
First, we note that
\begin{equation*}
W_C(X,Y)=\sum\limits_{c\in C}X^{n-w(c)} Y^{w(c)}=X^n \sum\limits_{c\in C}(\frac{Y}{X})^{w(c)},
\end{equation*}
thus,
\begin{equation*}
W_{C^{\bot}}(\cosh\beta,\sinh\beta)=(\cosh\beta)^n \sum\limits_{c\in C^{\bot}}(\tanh\beta)^{w(c)}
\end{equation*}
and
\begin{equation*}
W_{C}(\cosh\alpha,\sinh\alpha)=(\cosh\alpha)^n \sum\limits_{c\in C}(\tanh\alpha)^{w(c)}.
\end{equation*}
We only prove the following identity
\begin{equation*}
(\cosh\beta)^n \sum\limits_{c\in C^{\bot}}(\tanh\beta)^{w(c)}=\text{det}(A(C))(\frac{\sinh 2\beta}{2})^{\frac{n}{2}} (\cosh\alpha)^n \sum\limits_{c\in C}(\tanh\alpha)^{w(c)},
\end{equation*}
or equivalently,
\begin{equation*}
\sum\limits_{c\in C^{\bot}}(\tanh\beta)^{w(c)}=\text{det}(A(C))(\frac{\sinh 2\beta}{2})^{\frac{n}{2}} (\frac{\cosh\alpha}{\cosh\beta})^n \sum\limits_{c\in C}(\tanh\alpha)^{w(c)}.  \tag{4.6}
\end{equation*}
By the MacWilliams identity of $C$ ($q=2$, see (1.1))
\begin{equation*}
\sum\limits_{c\in C^{\bot}} z^{w(c)}=\frac{1}{|C|}(1+z)^n \sum\limits_{c\in C} (\frac{1-z}{1+z})^{w(c)}.
\end{equation*}
Let $z=\tanh\beta$, note that
\begin{equation*}
\frac{1-z}{1+z}=e^{-2\beta}=\tanh\alpha,
\end{equation*}
we thus have
\begin{equation*}
\sum\limits_{c\in C^{\bot}}(\tanh\beta)^{w(c)}=\frac{1}{|C|}(1+\tanh\beta)^n \sum\limits_{c\in C}(\tanh\alpha)^{w(c)}.  \tag{4.7}
\end{equation*}
By lemma 4.6, we have the following identity
\begin{equation*}
\text{det}(A(C))(\frac{\sinh 2\beta}{2})^{\frac{n}{2}} (\frac{\cosh\alpha}{\cosh\beta})^n=\frac{1}{|C|}(1+\tanh\beta)^n.  \tag{4.8}
\end{equation*}
Comparing the right hand sides of (4.6) and (4.7), we have (4.6) by (4.8), and thus we have (4.5). This is the proof of Theorem 3.\\
\hspace*{12cm} $\Box$

\section{Conclusion}

We show that in this paper, the statistical significance of MacWilliams theorem is to provide a real instance of the classical Gauss distribution for a code. If $C$ is a binary code, for example, $A(C)$ is the lattice associated with $C$, then the discrete Gauss measure over $A(C)$ is corresponding with the MacWilliams distribution over $C$. The further question is to give a precise lower bound for the smoothing parameter $\eta_{\epsilon}(C)$ of the arbitrary codes. Now we have a rough lower bound of it for linear codes as $\eta_{\epsilon}(C)\geqslant [(\frac{q^{n-k}}{1+\epsilon})^{\frac{1}{n}}-1]/(q-1)$. We will discuss this problem in the following works.

To generalize the results given by Theorem 3, it is natural to establish the MacWilliams type formula for the nu-function over arbitrary lattices. As we already pointed out that the Poisson formula for nu-function is missing, one of alternative method is to generalize the construction A, or other constructions for a lattice, so that we may transform this problem to a code over $F_q$, and make use of the MacWilliams identity. However, we don't have  any new results in this direction.

\end{document}